\documentclass{article}
\usepackage{spconf,amsmath,graphicx}
\usepackage{multirow}
\usepackage{tabularx}
\usepackage{xcolor}
\usepackage{booktabs}
\usepackage{array}

\usepackage{xurl}
\usepackage{etoolbox}
\usepackage{setspace}
\usepackage{balance}
\usepackage[symbol]{footmisc}

\newcolumntype{Y}{>{\centering\arraybackslash}X}           
\newcolumntype{R}{>{\raggedright\arraybackslash}p{0.28\textwidth}} 

\usepackage{amssymb}
\usepackage{microtype}
\usepackage{cite}
\usepackage{xcolor}
\usepackage{hyperref}

\title{In-Sync: Adaptation of Speech Aware Large Language Models for ASR with Word Level Timestamp Predictions}
\name{\shortstack{%
    Xulin Fan$^{1}$\sthanks{Work done during an internship at IBM Research}\qquad
    Vishal Sunder$^{2}$\qquad 
    Samuel Thomas$^{2}$\\
    Mark Hasegawa-Johnson$^{1}$\qquad
    Brian Kingsbury$^{2}$\qquad
    George Saon$^{2}$}}

\address{%
  $^{1}$University of Illinois Urbana-Champaign \\
  $^{2}$IBM Research
}
\begin{document}
\ninept
\maketitle
\begin{abstract}
Recent advances in speech–aware language models have coupled strong acoustic encoders with large language models, enabling systems that move beyond transcription to produce richer outputs. Among these, word-level timestamp prediction is critical for applications such as captioning, media search, and multimodal synchronization, yet it is often handled by external alignment tools. In this work, we extend an existing speech-aware language model to predict timestamps directly alongside transcripts. We introduce a set of novel lightweight training strategies that improve alignment robustness while preserving recognition quality. Experiments across multiple datasets show that these strategies not only enhance timestamp accuracy, but also yield gains in overall ASR performance. Together, they demonstrate an efficient and unified approach to speech recognition with precise timestamp prediction.
\end{abstract}
\section{Introduction}
\label{sec:intro}

The field of automatic speech recognition (ASR) has been fundamentally reshaped over the last decade, primarily through self-supervised learning (SSL). This paradigm began with pretrained acoustic encoders trained on massive unlabeled audio. Seminal models such as wav2vec 2.0~\cite{baevski2020wav2vec} learned representations directly from raw waveforms via contrastive masking, while HuBERT~\cite{hsu2021hubert} improved this approach using predictive losses over clustered discrete units. These frameworks proved highly effective at capturing acoustic and phonetic features, establishing pretrain–finetune pipelines that now form the backbone of modern ASR.

Building on these advances, recent work connects pretrained acoustic models with large language models (LLMs), giving rise to speech large language models (SpeechLLMs) that combine the perceptual strength of audio encoders with the linguistic knowledge of LLMs. AudioPaLM~\cite{rubenstein2023audiopalm} extends PaLM-2 to unify speech understanding and generation, SpeechGPT~\cite{zhang2023speechgpt} adapts LLMs for spoken dialogue, SALMONN~\cite{tang2023salmonn} augments LLaMA~\cite{touvron2023llama} with audio inputs to create general-purpose ``hearing models'', and SLAM-LLM~\cite{ma2024slamllm} describes a straightforward method for integrating an acoustic encoder with a LLM. This trend shifts ASR from transcription-only toward holistic spoken language processing and interactive applications.

Beyond accurate transcription, however, many practical applications require fine-grained temporal alignment between speech and text. Word-level timestamps are essential for applications such as closed captioning, video indexing, keyword-based audio retrieval, and multimodal synchronization. Traditionally, timestamps have been produced by forced alignment using hidden Markov models (HMMs), implemented in toolkits such as HTK \cite{young2002htk} and in hybrid ASR systems such as Kaldi \cite{povey2011kaldi}. The Montreal Forced Aligner \cite{mcauliffe2017montreal} makes such pipelines more accessible, but they still require separate alignment passes and additional acoustic models depending on the pronunciation dictionary. Similarly, The Nemo Forced Aligner (NFA) \cite{rastorgueva2023nfa} applies Viterbi decoding to the output of CTC-based models to derive timestamps. There are multiple works~\cite{chen2021e2etimestamp, bain2023whisperx} that builds upon a two-pass approach where either ground-truth transcript is given or we first estimate the transcript and then use the transcript and the audio to estimate the timestamps with a second pass. For example, WhisperX~\cite{bain2023whisperx} improves upon the popular Whisper ASR model~\cite{radford2023whisper} by adding an forced alignment module that refines timestamps at the word level, achieving good accuracy and efficiency. Another line of work~\cite{sunder2024pointer, zusag2024crisperwhisper} shows that speech–text alignment can be effectively learned using attention scores between speech and text modalities. Sunder et al. \cite{sunder2024pointer} propose a pointer network supervised by an ASR encoder–decoder model, alleviating the need for a pronunciation dictionary. CrisperWhisper~\cite{zusag2024crisperwhisper} proposes to leverage cross-attention score of a whisper model with modified tokenizer to learn accurate word-level timestamps. 
\begin{figure*}[ht]
  \centering
  \includegraphics[width=1\linewidth]{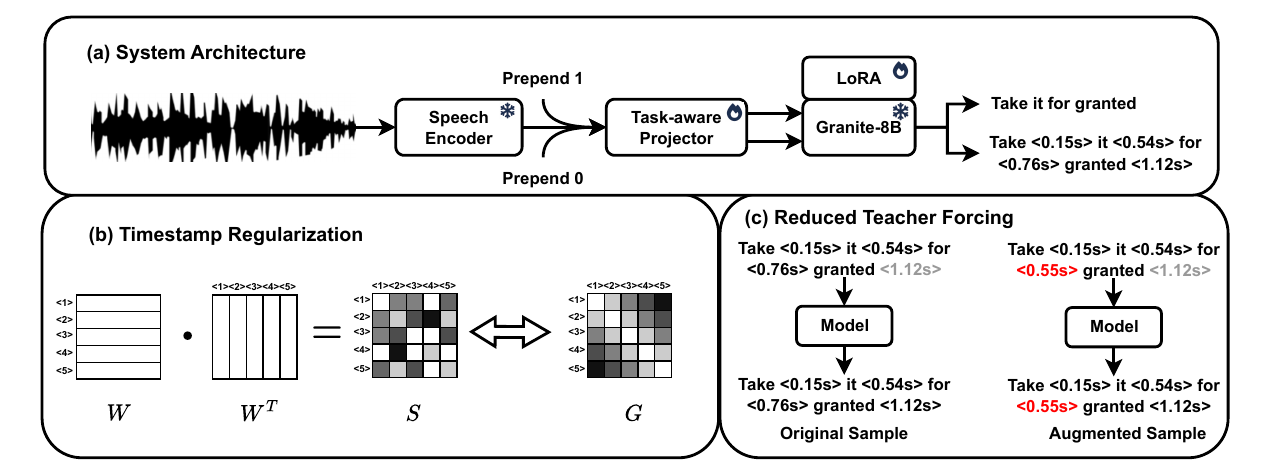}
  \vspace{-10pt}
  \caption{(a) Overall architecture of the proposed In-Sync framework for joint transcription and timestamp prediction using Granite-speech. (b) Diagram of timestamp embedding space regularization with $N{=}5$, where the similarity matrix $S$ is encouraged to match a structured target $G$. (c) Illustration of reduced teacher forcing during autoregressive generation, where a timestamp token is randomly corrupted to encourage model robustness.}
  \label{fig:signal_model}
  \vspace{-2pt}
\end{figure*}

Compared to traditional alignment approaches, end-to-end approaches that perform one-pass Speech Recognition With Timestamps (SRWT) directly 
have emerged as a promising alternative \cite{ nvcanarytimestamp, chu2023qwenaudio}. These models predict timestamp information alongside transcription, reducing reliance on external alignment tools or complicated model architecture design. However, such methods could introduce trade-offs, as timestamp prediction may compete with recognition accuracy or require architectural modifications that increase complexity.
Qwen-Audio~\cite{chu2023qwenaudio} integrates audio understanding with the Qwen LLM, enabling multimodal reasoning and fine-grained timestamp prediction as part of its broader capability set. These approaches highlight the opportunity to elevate timestamp prediction into a first-class objective within speech–language modeling frameworks. Moreover, one additional potential benefit is that the additional supervision from timestamps might foster better performance of automatic speech recognition and reduce hallucination.

We pursue this goal within Granite-speech~\cite{saon2025granitespeech}, a recently proposed Speech-aware LLM with strong ASR performance. We make extensions to the model that enable joint transcription and word-level timestamp prediction (In-Sync), eliminating the need for external aligners or costly post-processing. To stabilize timestamp training, we introduce three novel techniques designed for LLM adaptation:
\begin{enumerate}
\item \textbf{Speech Length Augmentation.} Concatenating consecutive utterances~\cite{asrconcat} balances the long-tail timestamp distribution and improves coverage of large timestamp tokens.
\item \textbf{Timestamp Embedding Regularization.} An auxiliary loss enforces structured similarity among timestamp embeddings, encouraging monotonic temporal progression.
\item \textbf{Reduced Teacher Forcing.} Randomly corrupting timestamp inputs mitigates over-reliance on ground-truth history, improving robustness in autoregressive generation.
\end{enumerate}

Together, these contributions enable effective timestamp prediction within Granite-speech, advancing toward end-to-end speech recognition with temporal grounding.

\begin{table*}[t]
\centering
\resizebox{\textwidth}{!}{%
\renewcommand{\arraystretch}{1.2}
\begin{tabularx}{\textwidth}{p{0.14\textwidth} l|*{9}{Y}}
\toprule
\textbf{Model} & \textbf{Metric} & AVG & LS-C$^{*}$ & LS-O$^{*}$ & CV$^{*}$ & AMI$^{*}$ & VOXP$^{*}$ & MLS$^{*\dagger}$ & TIMIT$^{\dagger}$ & BUCK$^{\dagger}$ \\
\midrule
\multicolumn{11}{l}{\textbf{External Baselines}} \\
\midrule
\multirow{2}{*}{\shortstack[l]{CrisperWhisper~\cite{zusag2024crisperwhisper}}\footnotemark}
  & WER & \textbf{5.60} & 1.71 & 3.72 & \textbf{7.76} & \textbf{8.43} & 6.03 & \textbf{5.26} & \textbf{1.29} & \textbf{10.63} \\
  & AAS & 53.65 & 30.20 & 33.84 & 119.37 & 64.80 & \textbf{54.93} & 48.36 & 34.30 & 43.41\\
  \midrule
\multirow{3}{*}{\shortstack[l]{Qwen-Audio~\cite{chu2023qwenaudio}}}
  & WER & 10.27 & 2.19 & 4.59 & 10.70\footnotemark & 31.82 & 7.22 & 7.54 & 5.96 & 12.15 \\
  & AAS & -- & 16.67 & 18.87 & -- & 55.64 & -- & \textbf{31.87} & 22.25 & \textbf{23.17} \\
  & MAL & -- & 0.57 & 0.54 & -- & 18.64 & -- & 4.51 & 0.06 & 0 \\
  \midrule

\multicolumn{11}{l}{\textbf{Granite-Speech ASR Baseline and Ablations with In-Sync}} \\
\midrule
ASR-only Baseline & WER & 6.55 & 1.72 & 3.68 & 8.95 & 9.95 & 6.31 & 6.84 & 2.85 & 12.09 \\
\midrule
\multirow{3}{*}{\shortstack[l]{Mixed Training}}
  & WER & 6.71 & 1.81 & 3.82 & 10.51 & 10.09 & 7.02 & 5.81 & 3.13 & \textbf{11.51} \\
  & AAS & 41.66 & 33.75 & 24.42 & 68.22 & \textbf{42.76} & 76.96 & 39.72 & 20.23 & 27.22 \\
    & MAL & 0.14 & 0.42 & 0.2 & 0.01 & 0 & 0.49 & 0 & 0 & 0 \\
\midrule
\multirow{3}{*}{\shortstack[l]{Mixed Training\\+ Length Aug}} 
  & WER & 6.60 & 1.72 & \textbf{3.65} & 8.97 & 10.69 & 6.05 & 5.68 & 4.18 & 11.84 \\
  & AAS & 41.41 & 13.37 & 17.36 & 56.03 & 46.17 & 116.37 & 35.13 & 20.48 & 26.38\\
      & MAL & 0.02 & 0 & 0 & 0.02 & 0 & 0.11 & 0.03 & 0 & 0 \\
\midrule
\multirow{3}{*}{\shortstack[l]{Mixed Training\\+Length Aug \\+ Timestamp Reg}}
  & WER & 6.34 & \textbf{1.62} & 3.69 & 9.40 & 9.79 & 6.15 & 5.69 & 2.53 & 11.88 \\
  & AAS & 37.23 & 12.61 & 16.55 & 68.70 & 43.48 & 73.72 & 34.72 & 20.22 & 27.81 \\
      & MAL & 0.08 & 0.08 & 0.17 & 0 & 0 & 0.33 & 0.03 & 0 & 0 \\
\midrule
\multirow{3}{*}{\shortstack[l]{Mixed Training\\+Length Aug \\+ Reduced TF}}
  & WER & 6.47 & 1.64 & \textbf{3.65} & 8.89 & 10.87 & \textbf{5.95} & 5.63 & 3.04 & 12.07 \\
  & AAS & \textbf{35.89} & \textbf{12.44} & \textbf{16.36} & \textbf{54.53} & 44.61 & 77.51 & 34.94 & \textbf{19.89} & 26.85 \\
        & MAL & 0.06 & 0.04 & 0.03 & 0.03 & 0 & 0.22 & 0.08 & 0 & 0.10 \\
\bottomrule
\end{tabularx}
}
\caption{Comparison across datasets for automatic speech recognition (ASR) and speech recognition with timestamps (SRWT). Word error rate (WER$\downarrow$) in percentage is used for evaluating ASR task performance, while accumulated averaging shift (AAS$\downarrow$)~\cite{shi20222e2timestamp2} in miliseconds (ms) and percentage of malformed samples (MAL$\downarrow$) measure SRWT task performance in terms of timestamp accuracy and stability to form a correct interleaved sequence. A dash (–) indicates that the model failed to follow the task prompt and instead hallucinated or performed a different task on that dataset. Datasets marked with $^{*}$ use timestamps obtained via the Montreal Forced Aligner, with samples failing alignment excluded. The TIMIT and Buckeye datasets have manual timestamp annotations. Datasets marked with $^{\dagger}$ are evaluated in a zero-shot setting for ASR-only baseline and all our In-Sync variants. The leftmost column with ``AVG'' denotes the average metrics across all datasets.}
\label{tab:results}
\end{table*}
\section{Method}
\label{sec:method}
In this section, we present the core components of In-Sync, as illustrated in Figure~\ref{fig:signal_model}. Section~\ref{sec:architecture} introduces the overall model architecture (Figure~\ref{fig:signal_model}a), which follows the Granite-speech-8B framework and comprises a pretrained audio encoder, a task-aware projector, and a large language model. Section~\ref{sec:training} outlines our multi-task training scheme that jointly optimizes for both ASR and speech recognition with word-level timestamps (SRWT) using task-specific prompts and a task-aware adapter. Section~\ref{sec:lengthaug} describes our speech length augmentation strategy, which improves coverage of long-range timestamp tokens by concatenating utterances. In Section~\ref{sec:embedreg}, we propose a timestamp embedding regularization loss (Figure~\ref{fig:signal_model}b) that aligns the learned timestamp similarity structure with a structured Gaussian prior. Finally, Section~\ref{sec:reducedtf} presents our reduced teacher forcing strategy (Figure~\ref{fig:signal_model}c), which mitigates timestamp error propagation by randomly corrupting timestamp tokens in the input during training.

\subsection{Model Architecture}
\label{sec:architecture}
We adopt a similar architecture to the Granite-speech-8B model~\cite{saon2025granitespeech} as the foundation for our experiments. The system comprises three components: a pretrained audio encoder, a speech adapter, and a pretrained large language model (LLM). Specifically, we use a 10-layer Conformer as the audio encoder~\cite{saon2025granitespeech}, a multi-layer perceptron (MLP) as the adapter~\cite{ma2024slamllm}, and Granite-3.3-8B-Instruct as the text LLM. As proposed in the original Granite-Speech framework, we freeze the speech encoder and the Granite LLM, while training the speech adaptor and the LoRA~\cite{hu2022lora} module applied to the LLM.

\subsection{Training Scheme}
\label{sec:training}
Following prior work such as Qwen-Audio~\cite{chu2023qwenaudio}, we formulate training as a multi-task learning problem involving both ASR and SRWT. During training, each input sample is randomly assigned to either the ASR or SRWT task with equal probability. The same speech input is paired with different text targets depending on the task, and task-specific prompts are prepended to the LLM input to condition its behavior appropriately.

To further improve task separation and training stability, we design the speech adapter to be task-aware. Concretely, a task indicator token is prepended to the speech embedding sequence as input to the speech adapter, allowing the adapter to generate distinct representations for ASR and SRWT, respectively.

For both tasks, targets are represented as text sequences, tokenized using the Granite tokenizer and trained with the next token prediction objective. For SRWT, we introduce one new token per 10ms interval, resulting in a total of 6000 additional tokens to cover the full 60-second maximum input duration supported by Granite-speech. Timestamp tokens are inserted into the transcript, producing interleaved word-timestamp sequence training targets.

\subsection{Speech Length Augmentation}
\label{sec:lengthaug}
Due to the nature of word-level timestamp prediction, 
timestamp tokens in the training data follow a heavy-tailed distribution: shorter timestamps are heavily represented, while larger timestamps are much rarer. This skewed distribution biases the model towards predicting earlier timestamps and impairs its ability to generalize to longer-duration speech segments. Prior work has explored simple sample concatenation as a data augmentation strategy to enhance ASR robustness~\cite{asrconcat}. Inspired by these findings, we apply a similar augmentation technique by concatenating pairs of utterances during training and shifting the timestamp targets of the second utterance by the duration of the first. This augmentation effectively extends the timestamp range covered in training, helping balance the timestamp token distribution and enabling more accurate prediction of larger timestamps.

\footnotetext[1]{23 CV, 36 VOX, 3 MLS samples excluded from CrisperWhisper evaluation due to decoding error during inference.}
\footnotetext[2]{Undergoes a rule-based post-processing for correcting a fixed error pattern before evaluation.}

\subsection{Timestamp Embedding Regularization}
\label{sec:embedreg}
Timestamp tokens are inherently ordered and monotonically increasing, reflecting the progression of time. However, the standard next-token prediction objective used in language modeling does not explicitly enforce or leverage this structure. As a result, especially under limited data conditions, the model may struggle to learn a coherent geometric organization of the timestamp embeddings.

To address this issue, we propose an auxiliary timestamp embedding regularization loss that encourages the learned embeddings of timestamp tokens to maintain a smooth and ordered topology. Let $W \in \mathbb{R}^{N \times d}$ denote the embedding matrix corresponding to the $N$ timestamp tokens, where each row of $W$ is normalized to unit norm. We define a cosine similarity matrix $S = W W^\top \in \mathbb{R}^{N \times N}$, where $S_{ij}$ measures the similarity between the $i$-th and $j$-th timestamp embeddings.

We construct a target similarity matrix $G \in \mathbb{R}^{N \times N}$ using a Gaussian kernel centered along the diagonal:

\[
G_{ij} = \exp\left(-\frac{(i - j)^2}{2\sigma^2}\right)
\]
for a fixed standard deviation $\sigma$. The regularization loss is then defined as the mean squared error between the predicted and target similarity matrices:

\[
\mathcal{L}_{\text{reg}} = \frac{1}{N^2} \sum_{i=1}^{N} \sum_{j=1}^{N} (S_{ij} - G_{ij})^2
\]

This loss promotes timestamp embeddings whose cosine similarities reflect their temporal ordering, with high similarity between neighboring tokens and decreasing similarity as timestamps diverge. During training, $\mathcal{L}_{\text{reg}}$ is added to the standard next-token prediction loss with a tunable weight $w_{reg}$.

\subsection{Reduced Teacher Forcing}
\label{sec:reducedtf}
One common error pattern we observe in the SRWT task when using large language models is the propagation of timestamp errors. Due to the autoregressive generation and the default teacher forcing training scheme, timestamp tokens in the input are always ground-truth during training. This setup implicitly encourages the model to rely heavily on relative offsets which simply predict the current timestamp based on the previous one. While this strategy may work in ideal cases, it leads to cascading failures: an error in a single timestamp prediction propagates forward, causing subsequent timestamps to be misaligned even if the model has correctly learned local durations. 

To address this issue, we propose a reduced teacher forcing strategy that limits reliance on prior timestamps. During training, timestamps in the input sequence are randomly corrupted with smaller values at probability $p$, encouraging the model to balance global alignment (absolute word position) with local dependence on preceding timestamps. By relaxing the assumption of perfect past timestamps, the model learns more robust and generalizable alignment at inference.

\section{Experiments}
\label{sec:experiment}

\subsection{Experiment Configuration}
We train our models on four datasets—LibriSpeech~\cite{panayotov2015librispeech}, CommonVoice~\cite{ardila2019commonvoice}, AMI-IHM~\cite{kraaij2005ami}, and VoxPopuli~\cite{wang2021voxpopuli}—and evaluate on eight datasets: LibriSpeech test-clean (LS-C), LibriSpeech test-other (LS-O), CommonVoice (CV), AMI-IHM (AMI), VoxPopuli (VOXP), MLS English (MLS)~\cite{pratap2020mls}, TIMIT~\cite{garofolo1993timit}, and Buckeye (BUCK)~\cite{pitt2005buckeye}. For datasets lacking timestamp annotations, we apply the Montreal Forced Aligner (MFA)~\cite{mcauliffe2017montreal}, using a higher beam size for training data and a lower beam size for test data to ensure high-quality alignment for evaluation. To maintain consistency across human-annotated and MFA-aligned datasets and to reduce the output sequence length, we always set the start timestamp of each word to the end timestamp of the preceding word so the language model only needs to output one timestamp for each word. All models are trained for 400k steps with the AdamW optimizer, a peak learning rate of 0.0001, and a 1000 steps warm-up schedule. The speech adaptor always has a temporal downsampling rate of 5. The Granite LLM is trained with LoRA targeting the query and value projections with a rank of 32 and an alpha of 64. We use a batch size of 4 per GPU across 4 GPUs.

For data augmentation, we construct a length-augmented version of LibriSpeech by concatenating consecutive sample pairs into longer utterances. Timestamp regularization introduces a Gaussian prior with standard deviation $\sigma = N/4$ and a loss weight of $w_{\text{reg}} = 0.1$. Reduced teacher forcing is applied with probability $p = 0.2$ by randomly replacing each timestamp token in the input sequence with a smaller token uniformly sampled between the first timestamp and the ground-truth current timestamp.

\subsection{Results}
\label{sec:results}
We present our results in Table~\ref{tab:results}, including two external-baselines~\cite{chu2023qwenaudio, zusag2024crisperwhisper}\footnote{\url{https://huggingface.co/Qwen/Qwen-Audio}}, an ASR-only baseline trained with the same Granite-Speech architecture, and several ablations of our proposed In-Sync framework for ASR and SRWT. We take the predicted end time of each word for evaluation with the SRWT metrics. For SRWT inference, a small number of samples produce malformed sequences with mismatched counts of word and timestamp tokens. Since alignment cannot be computed in these cases, they are excluded from AAS evaluation and the percentage of malformed samples are reported as ``MAL''. 

As shown in Table~\ref{tab:results}, when shifting to mixed training of ASR and SRWT, we see timestamp supervision enables reasonable alignment accuracy, but comes at the cost of degraded ASR performance compared to the ASR-only baseline.

Length augmentation improves timestamp accuracy on datasets with longer utterances, showing the benefit of exposing the model to extended temporal contexts. However, performance degrades on some datasets, suggesting the augmentation introduces distributional mismatch when utterances are naturally short. Timestamp regularization proves most effective for balancing the two objectives, achieving the better WER while also reducing AAS. Reduced teacher forcing attains the strongest overall timestamp accuracy, enhancing alignment robustness across most datasets. It also narrows the WER gap relative to the baseline, indicating that controlled noise during training improves the robustness of autoregressive generation at inference.

For comparison with external baselines~\cite{chu2023qwenaudio, zusag2024crisperwhisper}, we observe that while Qwen-Audio predicts timestamps with reasonable accuracy on clean speech datasets, it fails to follow the SRWT prompt and does not generate timestamps on most samples of CommonVoice and VoxPopuli, so these SRWT evaluations are marked with “–”. CrisperWhisper achieves best word error rate on most datasets evaluated but it's worth noting that it's initialized with the whisper-large-v2~\cite{radford2023whisper} which has been trained on significantly more data. Our systems handle diverse datasets with varying noise conditions and recording environments, achieving a better average WER than Qwen-audio and a better average AAS score over CrisperWhisper. On TIMIT and Buckeye, our model operates in a purely zero-shot setting, and the mismatch between MFA-aligned training labels and human-annotated test labels introduces a domain gap that limits performance.

\subsection{Limitations}
We note two limitations in our current framwork to guide future work on word-level timestamps with speech-aware language models. First, while timestamp regularization and reduced teacher forcing each improve robustness, they combine poorly: the corruption in reduced teacher forcing breaks the monotonic structure that the regularization seeks to enforce. Second, predicting only end-of-word timestamps shortens targets and reduces malformed sequences but prevents explicit modeling of silence without post-processing. Introducing a dedicated silence token would avoid start–end pairs, yet in our tests such unseen word tokens degraded performance. We leave both directions to future work.

\section{Conclusion}
\label{sec:conclusion}

In this paper, we extend the Granite-speech framework to support joint ASR and word-level timestamp prediction. While naive multitask training yields reasonable timestamp accuracy, it degrades recognition quality. To mitigate this trade-off, we introduce auxiliary strategies including length augmentation, timestamp embedding regularization, and reduced teacher forcing, to strengthen timestamp accuracy without harming ASR. Our results show that  Granite-speech can be effectively adapted for unified transcription and precise temporal alignment, enabling applications that demand both high transcription quality and precise temporal alignment.


 \vfill\pagebreak

%
\bibliographystyle{IEEEbib}
\bibliography{strings,refs}

\end{document}